\documentstyle[12pt]{article}

\newcommand{\matel}[3]{\langle #1|#2|#3\rangle}
\newcommand{\ra}{\rightarrow}

\newcommand{\sG}{\sigma \cdot G}

\newcommand{\aver}[1]{\langle #1\rangle}

\newlength{\dinwidth}
\newlength{\dinmargin}
\begin{document}
{}~~\\
\vspace{1cm}
\begin{flushright}
UND-HEP-95-BIG09\\
August 1995\\
\end{flushright}

\begin{center}
\begin{Large}
\begin{bf}
%
%
INCLUSIVE $B_c$ DECAYS AS A QCD LAB

\end{bf}
\end{Large}
\vspace{5mm}
\begin{large}
%
%
I.I. Bigi\\
\end{large}
%
%
Physics Dept., University of Notre Dame du Lac,
Notre Dame, IN 46556, U.S.A.\\
e-mail address: BIGI@UNDHEP.HEP.ND.EDU
\vspace{5mm}
\end{center}
\noindent
\begin{abstract}
Phenomenological models of heavy flavour decays differ
significantly in their predictions of global features of
$B_c$ decays, like the $B_c$ lifetime or the relative
weight of $c\ra s$ and $b\ra c$ transitions.
The $1/m_Q$ expansion which is directly based on
QCD allows predictions on
the pattern to be expected, namely $\tau (B_c)$ to lie well
below 1 psec with $c\ra s$ dominating over $b\ra c$ and a reduced
semileptonic branching ratio.
Due to interference effects one also predicts a
lower charm content in the final states of $B_c$
decays than naively anticipated. The numerical aspect of the
predictions, however, has to be viewed with considerable
caution since one cannot expect the $1/m_c$ expansion to
converge readily for $\Delta C=1$ transitions.

\end{abstract}

\section{Introduction}

$B_c$ mesons consisting of two heavy quarks --
$B_c = (b\bar c)$ -- are not easily produced. On the
other hand it is highly desirable to obtain large samples of them.
For their study would deepen our quantitative understanding
of the inner workings of QCD in a significant way: one expects a
rich spectroscopy for the $(b\bar c)$ boundstates probing the
inter-quark potential at distances intermediate to those
determining quarkonia spectroscopy in the charm and beauty
systems \cite{QUIGG}; the Isgur-Wise function for the striking channel
$B_c \ra l \nu \psi$ can be calculated; the weak $B_c$ decays
reflect a multi-faceted interplay of various dynamical
mechanisms. It is this last aspect I will analyze in this note: in
Sect.2 I will review previous phenomenological descriptions of
$B_c$ decays; in Sect.3 I introduce a genuine
QCD treatment based on the heavy quark expansion and apply it
to inclusive $B_c$ transitions; in Sect.4 I discuss the final states
in $B_c$ decays before presenting my conclusions in
Sect.5.

\section{A First Phenomenological Look at $B_c$ Decays}

There are three classes of transitions contributing to $B_c$ decays
with roughly comparable strength; they appear to be easily
distinguishable on the diagrammatic level.
The first two are the decay of the $b$ quark and that of the
$\bar c$ antiquark. Since $b\ra c$ and $\bar c\ra \bar s$
transitions do not
interfere with each other in any appreciable way, one can
cleanly separate their widths. The only subtlety here is
that  $b\ra c \bar cs$  decays
lead to two $\bar c$ antiquarks in
$B_c$ decays; there arises then an interference
between different decay amplitudes that is usually
referred to as PI. The third class of transitions is produced by
Weak Annihilation (WA) of $b$ with $\bar c$. To lowest order in
the strong interactions the WA amplitude suffers helicity and
wavefunction suppression (the latter reflecting the practically
zero range of the weak interactions). Yet for
$B_c \ra s\bar c$ they are represented by $m_c/m_b$ and
$f(B_c)/m_b$ and thus relatively mild ($f(B_c) \sim 450-700$
MeV \cite{QUIGG});
furthermore these reductions are partially offset by the factor
$16\pi ^2$ reflecting the enhancement of two-body phase
space -- relevant for WA -- over three-body phase space
appropriate for the spectator decay.
As explained later, interference between $b$ decay and WA
can arise; yet this is usually ignored in phenomenological
analyses.
One thus writes down for the nonleptonic width:
$$\Gamma _{NL}(B_c)=
\Gamma _{\bar c\ra \bar s d\bar u}^{spect}(B_c) +
\Gamma _{b\ra c\bar u d}^{spect}(B_c)  +
\Gamma _{b\ra c\bar c s}^{spect}(B_c)  -
|\Delta \Gamma _{b\ra c \bar cs}^{PI}(B_c)|
+ \Gamma _{WA}(B_c)
\eqno(1)$$
The expression simplifies for the semileptonic width since no
interference occurs there and WA does not contribute
(at least to lowest order in the strong coupling)
\footnote{Only $B_c \ra \tau ^+ \nu $ leading to a purely
leptonic final state possesses an appreciable rate.}:
$$\Gamma _{SL}(B_c)=
\Gamma _{\bar c\ra \bar s l\nu}^{spect}(B_c) +
\Gamma _{b\ra c l \nu}^{spect}(B_c)  \; .
\eqno(2)$$
Very naively one might equate
$\Gamma _{\bar c \ra \bar s}(B_c)$ with $\Gamma (D^0)$
and $\Gamma _{b \ra c}(B_c)$ with $\Gamma (B_d)$
and thus
$$\Gamma (B_c) \simeq \Gamma (D^0) + \Gamma (B_d) \; .
\eqno(3)$$
If eq.(3) were to hold, the $B_c$ lifetime would be rather
short, namely
$$ \tau (B_c) \sim 3\cdot 10^{-13} \; sec \eqno(4)$$
and $B_c$ decays would be dominated by $\bar c\ra \bar s$
over $b\ra c$ in the ratio of roughly  4:1. It is quite natural,
though, to suspect that eq.(3) represents a gross
oversimplification. Two specific alternatives have been
suggested:

\noindent (i) The phase space in $B_c = (b\bar c) \ra
b \bar s d \bar u \simeq B_s (d \bar u)$ is more limited
than in $\bar D = (q \bar c) \ra q\bar s d \bar u \simeq
K (d \bar u)$. This could -- due to the high sensitivity of
the $c$ decay width to the available phase space -- reduce
$\Gamma _{\bar c\ra \bar s}(B_c)$ significantly relative
to $\Gamma (D)$. No such reduction is expected for
$\Gamma _{b\ra c}(B_c)$. Therefore the relative weight of $b\ra c$
transitions in $B_c$ decays gets enhanced. Using a simple
recipe for estimating the phase space dependance of the
quark decay width the authors of ref.\cite{LUSIGNOLI}
estimate
$$ \tau (B_c) \sim 5\cdot 10^{-13}\; sec \eqno(5)$$
with -- and that is the major difference to the naive
guestimate given above -- $\bar c \ra \bar s$
transitions now holding only
a slight edge over $b\ra c$ decays.

\noindent (ii) It has been advocated by Eichten and Quigg
that in the expression for the decay width --
$\Gamma _Q \propto G_F^2 m_Q^5$ -- one should use a quark mass
{\em reduced} by the binding energy inside
the hadron. For the ordinary
mesons $B_d$, $B_u$ and $B_s$ this can be seen effectively as
a redefinition of the quark mass. Yet for the more tightly
bound system $B_c$ there arises an observable difference:
the binding energy $\mu _{BE}$ being the same
for the charm and for the beauty mass
represents a larger fraction of the charm mass than of the
beauty mass: $(m_c -\mu _{BE})/m_c = 1- \mu _{BE}/m_c <
(m_b -\mu _{BE})/m_b = 1- \mu _{BE}/m_b$.
Inside the $B_c$ meson
the $\bar c \ra \bar s$ rate will therefore
be more reduced than
the $b\ra c$ rate.  Furthermore
the impact of this mass shift on the
width is greatly enhanced:
$\Delta \Gamma _Q/\Gamma _Q \simeq
5\cdot \mu _{BE}/m_Q$ due to $\Gamma _Q \propto m_Q^5$!
For $\mu _{BE}= 500$ MeV one finds
$\Gamma _{\bar c\ra \bar s}$ and
$\Gamma _{b\ra c}$ reduced by a factor of 6 and 1.7,
respectively! This leads to the guestimate
$$ \tau (B_c) \sim 1.3 \; psec \; , \eqno(6)$$
i.e., a considerably longer $B_c$ lifetime; the
$b\ra c$ transitions now occur somewhat more frequently
than the $\bar c\ra \bar s$ ones.

The two questions raised above -- (i) whether the $B_c$
lifetimes is short, i.e. well below 1 psec, or `long', i.e.
roughly 1 psec or longer, and (ii) whether $B_c$ decays
are driven mainly by $b\ra c$ or by $\bar c\ra \bar s$
transitions -- are highly important and deserve study with
the best available theoretical tool, the $1/m_Q$ expansion.

\section{Treating $B_c$ Decays through a $1/m_Q$
Expansion}
\subsection{General Methodology}

In analogy to the treatment of
$e^+e^-\rightarrow hadrons$ one describes the transition
rate into an
inclusive final state $f$ through the imaginary part of a
forward scattering operator evaluated to second order in the weak
interactions \cite{SV,BUV}:
$$\hat T(Q\rightarrow f\rightarrow Q)=
i \, Im\, \int d^4x\{ {\cal L}_W(x){\cal L}_W^{\dagger}(0)\}
_T\eqno(7)$$
where $\{ .\} _T$ denotes the time ordered product and
${\cal L}_W$ the relevant effective weak Lagrangian expressed on
the
parton level. If the energy release in the decay is sufficiently large
one can express the {\em non-local} operator product in eq.(7) as an
infinite sum of {\em local} operators $O_i$ of increasing dimension
with
coefficients
containing higher and higher powers of $1/m_Q$.
The width for $H_Q\rightarrow f$ is obtained by
taking the
expectation value of $\hat T$ between the state $H_Q$.
For semileptonic and nonleptonic
decays treated through order $1/m_Q^3$ one arrives at the
following generic expression\cite{BUV}:
$$\Gamma (H_Q\ra f)=\frac{G_F^2m_Q^5}{192\pi ^3}|KM|^2
\left[ c_3^f\matel{H_Q}{\bar QQ}{H_Q}+
c_5^f\frac{
\matel{H_Q}{\bar Qi\sG Q}{H_Q}}{m_Q^2}+ \right.
$$
$$\left. +\sum _i c_{6,i}^f\frac{\matel{H_Q}
{(\bar Q\Gamma _iq)(\bar q\Gamma _iQ)}{H_Q}}
{m_Q^3} + {\cal O}(1/m_Q^4)\right]  \eqno(8)$$
where the dimensionless coefficients $c_i^f$ depend on the
parton level
characteristics of $f$ (such as the ratios of the final-state quark
masses
to $m_Q$); $KM$ denotes the appropriate combination of KM
parameters,
and $\sG = \sigma _{\mu \nu}G_{\mu \nu}$
with $G_{\mu \nu}$ being the gluonic field strength tensor.
The last term in eq.(8)
implies also the summation over the four-fermion operators with
different light flavours $q$.
It is through the quantities
$\matel{H_Q}{O_i}{H_Q}$ that the dependence on the
{\em decaying hadron} $H_Q$, and
on
non-perturbative forces in general, enters; they reflect the
fact that the weak decay of the heavy quark $Q$ does not
proceed
in empty space, but within a cloud of light degrees of
freedom -- (anti)quarks and gluons -- with which $Q$ and
its decay products can interact strongly.
These are matrix
elements
for on-shell hadrons $H_Q$; $\Gamma (H_Q\ra f)$ is thus
expanded into a power series in $\mu _{had}/m_Q < 1$. For
$m_Q\ra \infty$ the contribution from the lowest dimensional
operator obviously dominates; here it is the
{\em dimension-three}
operator $\bar QQ$. Since
$\matel{H_Q}{\bar QQ}{H_Q}=1+
{\cal O}(1/m_Q^2)$ holds, one reads off from eq.(8) that
the leading contribution to the total decay
width is
{\em universal} for all hadrons
of a given heavy-flavour quantum number;
i.e., for $m_Q\ra \infty$ one has derived -- from QCD proper --
the spectator picture.
Contributions from what is referred to as
WA and PI in the
original phenomenological descriptions are systematically and
consistently included through the dimension-six four-fermion
operators in eq.(8).

Yet the $1/m_Q$  expansion
goes well beyond reproducing familiar results. It shows the
leading nonperturbative corrections
to integrated inclusive rates to arise in order
$1/m_Q^2$ controlled by the expectation
values of dimension-five operators \cite{BUV}. These
terms had
been overlooked before.
What is crucial for our subsequent
analysis is the {\em absence} of contributions of order $1/m_Q$.
This is due to the fact that there is {\em no relevant
dimension-four} operator that cannot be removed by
applying the equation of motion \cite{CGG,BUV};
it can also be understood as due to a
subtle intervention of the local colour gauge
symmetry. A phenomenological ansatz on the other hand
where the quark mass appearing in the decay width
is reduced by a `binding energy' leads to large corrections
of order $1/m_Q$; see the discussion above eq.(6). This is
in clear conflict with what holds in QCD!

Using the equation of motion one can obtain a $1/m_Q$
expansion also for
the leading operator in eq.(8), $\bar QQ$. Its expectation value
can then be expressed as follows
\footnote{I use the relativistic normalization for the states.}:
$$\matel{H_Q}{\bar QQ}{H_Q} = 1-
\frac{\aver{(\vec p_Q)^2}_{H_Q}}{2m_Q^2}+
\frac{\aver{\mu _G^2}_{H_Q}}{2m_Q^2} + {\cal O}(1/m_Q^3)
\eqno(9)$$
where $\aver{(\vec p_Q)^2}_{H_Q}\equiv
\matel{H_Q}{\bar Q(i\vec D)^2Q}{H_Q}$ denotes the average
kinetic energy of the quark $Q$ moving inside the hadron
$H_Q$ and $\aver{\mu _G^2}_{H_Q}\equiv
\matel{H_Q}{\bar Q\frac{i}{2}\sG Q}{H_Q}$.

Eqs.(8,9) show that the nonperturbative contributions to the width
through order $1/m_Q^3$ can be expressed through
three expectation values: $\aver{\mu _G^2}_{H_Q}$,
$\aver{(\vec p_Q)^2}_{H_Q}$ and
$\matel{H_Q(p)}{\bar Q_L \gamma _{\mu}q_L)
(\bar q_L \gamma _{\nu}Q_L)}{H_Q(p)}$. The size of
the mesonic matrix element of the chromomagnetic operator
is obtained from the hyper-fine splitting:
$$ \aver{\mu _G^2}_{P_Q}\simeq
\frac{3}{4} (M_{V_Q}^2-M_{P_Q}^2) \; , \eqno(10)$$
where $P_Q$ and $V_Q$ denote the pseudoscalar and vector mesons,
respectively. For the average kinetic energy we have the
model-independant bound \cite{OPTICAL}
$$\aver{(\vec p_Q)^2}_{H_Q} \geq \aver{\mu _G^2}_{H_Q}
\eqno(11)$$
and we know it cannot be much larger than that.
The $1/m_Q^2$ corrections thus largely cancel in the
expectation value of the operator $\bar QQ$ between
{\em meson} states:
$$ \matel{P_Q}{\bar QQ}{P_Q}\simeq
1+{\cal O}(1/m_Q^3)$$
The expectation values for the four-quark operators
taken between meson states can be expressed in terms of a
single quantity, namely the decay constant:
$$\matel{H_Q(p)}{\bar Q_L \gamma _{\mu}q_L)
(\bar q_L \gamma _{\nu}Q_L)}{H_Q(p)}\simeq
\frac{1}{4} f^2_{H_Q}p_{\mu}p_{\nu} \eqno(12)$$
where factorization has been assumed.

\subsection{$B_c$ Decays through Order $1/m_Q^2$}

Since
$b\ra c$ and $\bar c \ra \bar s$ decays
do no interfere with each other in any practical way,
one can cleanly separate their widths:
$$\Gamma (B_c) = \Gamma _{b\ra c}^{decay}(B_c) +
\Gamma _{\bar c \ra \bar s}^{decay}(B_c) +
{\cal O}(1/m_{b,c}^3)  \eqno(13)$$
These widths are denoted as $\Gamma ^{decay}$ rather
than $\Gamma ^{spect}$ for a reason: they describe the
quark decays as proceeding in an environment shaped by
the other components of the decaying hadron $H_Q$ as expressed
by the expectation values $\aver{(\vec p_Q)^2}_{H_Q}$ and
$\aver{\mu _G^2}_{H_Q}$; thus they go beyond the
simple spectator picture.

The decay widths include
$1/m_Q^2$ corrections which
consist of the semileptonic and
nonleptonic components:
$$\Gamma _{b\ra cl\nu}^{decay}(B_c) =
\Gamma _0^{(b)} \cdot
\matel{B_c}{\bar bb}{B_c}
\left[ I_0(x_c,0,0) +
\frac{\aver{\mu _G^2}_{B_c}}{m_b^2}
(x\frac{d}{dx}-2)I_0(x_c,0,0)\right] \; , \eqno(14a)$$
$$\Gamma _{b\ra c q\bar q'}^{decay}(B_c) =
\Gamma _0^{(b)} \cdot N_C\cdot
\matel{B_c}{\bar bb}{B_c}
\left[ A_0[\sum I_0(x_c) +
\frac{\aver{\mu _G^2}_{B_c}}{m_b^2}
(x\frac{d}{dx}-2)\sum I_0(x_c)] - \right.$$
$$\left. 8A_2 \frac{\aver{\mu _G^2}_{B_c}}{m_b^2} \cdot
[I_2(x_c,0,0,) + I_2(x_c,x_c,0)]\right] \; . \eqno(14b)$$
$$\Gamma _0^{(b)} \equiv
\frac{G_F^2m_b^5}{192 \pi ^3}|V(cb)|^2 \eqno(14c)$$
where the following notations have been used:
$I_0$ and $I_2$ are phase-space factors:
$$I_0(x,0,0)= (1-x^2)(1-8x+x^2)-12x^2\log  x \eqno(15a)$$
$$I_2(x,0,0)=(1-x)^3\; , \; \; x_c=(m_c/m_b)^2 \eqno(15b)$$
$$I_0(x,x,0)=v(1-14x-2x^2-12x^3) + 24x^2(1-x^2)
\log \frac{1+v}{1-v}\, , \, v=\sqrt{1-4x}\eqno(15c)$$
$$I_2(x,x,0)=v(1+\frac{x}{2}+3x^2)-3x(1-2x^2)
\log \frac{1+v}{1-v} \, , \eqno(15d)$$
with $I_{0,2}(x,x,0)$ describing the $b\ra c \bar cs$
transition, and $\sum I_0(x)\equiv
I_0(x,0,0)+ I_0(x,x,0)$; $A_0= \eta J$,
$A_2=(c_+^2-c_-^2)/6$, where
$\eta = (c_-^2+2c_+^2)/3$, and $J$ represents the effect
of the subleading logarithms \cite{PETRARCA}.
With $x_c\simeq 0.08$ one obtains
$$I_0(x,0,0)|_{x=0.08}\simeq 0.56 \; \; , \; \;
I_2(x,0,0)|_{x=0.08}\simeq 0.78   \; \; for \; \;
b\ra c \bar ud$$
$$I_0(x,x,0)|_{x=0.08}\simeq 0.24 \; \; , \; \;
I_2(x,x,0)|_{x=0.08}\simeq 0.32 \; \; for \; \;
b\ra c \bar cs \; .$$
Since these functions are normalized to unity for $x=0$, one
notes that the final-state quark masses reduce the available
phase space quite considerably in this reaction.
The expressions are simpler for $c\ra s$:
$$\Gamma _{c\ra sl\nu}^{decay}(B_c) =
\Gamma _0^{(c)} \cdot
\matel{B_c}{\bar cc}{B_c}
\left[ I_0(x_s,0,0) +
\frac{\aver{\mu _G^2}_{B_c}}{m_c^2}
(x\frac{d}{dx}-2)I_0(x_s,0,0)\right] \; , \eqno(16a)$$
$$\Gamma _{c\ra s u\bar d}^{decay}(B_c) =
\Gamma _0^{(c)} \cdot N_C\cdot
\matel{B_c}{\bar cc}{B_c}
\left[A_0[I_0(x_s,0,0) +
\frac{\aver{\mu _G^2}_{B_c}}{m_c^2}
(x\frac{d}{dx}-2)I_0(x_s,0,0)] - \right.$$
$$\left. 8A_2 \frac{\aver{\mu _G^2}_{B_c}}{m_c^2}
\cdot
I_2(x_s,0,0,)\right]\; . \eqno(16b)$$
$$\Gamma _0^{(c)} \equiv
\frac{G_F^2m_c^5}{192 \pi ^3}|V(cs)|^2 \; , \; \;
x_s=\frac{m_s^2}{m_c^2}\eqno(16c)$$
and the radiative corrections lumped into $A_0$
and $A_2$ are given by the appropriate values for
$c_+$ and $c_-$. With $x_s
\sim 0.012$ one finds:
$$I_0(x,0,0)|_{x=0.012}\simeq 0.91 \; \; , \; \;
I_2(x,0,0)|_{x=0.012}\simeq 0.96 \; , $$
i.e. there is much less phase space suppression than for
$b\ra c$ transitions.

The transition {\em operators}
driving $B_c$ decays are the same
that generate $B$ and $D$ decays. However their
{\em expectation values} are evaluated for the $B_c$ state, rather
than the $B$ and $D$ state reflecting that the $b\ra c$ and
$\bar c\ra \bar s$ transitions proceed in a
different environment.
The expectation value of the chromomagnetic
operator is again given by hyperfine splitting between the
masses of $B_c^*$ and $B_c$. Those have not been measured
yet; on the other hand the theoretical predictions should
be quite reliable for those. With
$M(B_c^*)\simeq 6.33$ GeV and $M(B_c)\simeq 6.25$ GeV one
obtains
$$\matel{B_c}{\bar b \frac{i}{2}\sG b}{B_c} \simeq
 \matel{\bar B_c}{\bar c \frac{i}{2}\sG c}{\bar B_c}
\simeq 0.75\; (GeV)^2 \; , \eqno(17)$$
which is twice the value as for mesons with light antiquarks:
$$\matel{B}{\bar b \frac{i}{2}\sG b}{B} \simeq
0.37 \; (GeV)^2 \; \;  , \; \;
\matel{D}{\bar c \frac{i}{2}\sG c}{D} \simeq
0.41 \; (GeV)^2 \;  . \eqno(18)$$
Thus
$$\frac{\matel{B_c}{\bar b \frac{i}{2}\sG b}{B_c}}{m_b^2}
\simeq 0.033 \eqno(19a)$$
$$\frac{\matel{\bar B_c}{\bar c \frac{i}{2}\sG c}{\bar B_c}}
{m_c^2} \simeq 0.38 \; ; \eqno(19b)$$
i.e., this correction becomes quite large in the $c\ra s$
transition.

Putting everything together one finds for the $B_c$
width through order $1/m^2_{b,c}$:
$$ \Gamma (B_c) \simeq 0.95 \cdot \Gamma (B_d)
+ 0.75 \cdot \Gamma (D^0)
+{\cal O}(1/m^3_{b,c}) \sim (4.1 \cdot 10^{-13}
\; sec)^{-1} \eqno(20)$$
$$\frac{\Gamma _{b\ra c}(B_c)}{\Gamma (B_c)}
\sim 0.26 \; ; \eqno(21)$$
i.e., a short lifetime with $c\ra s$ transitions dominating
all $B_c$ decays! One also obtains a rather low
semileptonic branching ratio
$$BR_{SL}(B_c) \sim 6\; \% \eqno(22)$$
with {\em half} of the semileptonic $B_c$ decays being
generated by $b\ra c l\nu$. However these numbers
have to be taken with quite a grain of salt.
For the nonperturbative corrections in the
$c\ra s$ component of the $B_c$ width are very large,
as indicated by eq.(19b). Thus one can hope only for a
semi-quantitative treatment of that component.

\subsection{Order $1/m_Q^3$ Contributions}

In order $1/m_Q^3$ the explicitely flavour dependant terms
appear that had been anticipated in previous phenomenological
studies. Due to the large value predicted for $f(B_c)$ they
are quite sizable: as indicated in eq.(1)
PI reduces the rate for $b\ra c \bar cs$
to proceed inside $B_c$ mesons by $\sim$ 20 - 40\% and
WA contributes in an only mildly
suppressed manner.  In addition a more subtle effect
arises that had not been incorporated into eq.(1):
$\Gamma _{WA}(B_c)$ and
$\Gamma _{b\ra c}^{spect}(B_c)$ can no longer
be separated in a strict manner. For those two classes of reactions
-- for $b\ra c \bar ud$ as well as for $b\ra c \bar cs$
modes -- can {\em interfere} with each other once gluon
emission generates a $c \bar c$ pair in WA:
$B_c \ra b\bar c g^*g^* \ra s\bar c (c\bar c)_{g^*g^*}$. While this
observation has no relevant impact on the predicted
overall lifetime \cite{MIRAGE}, it becomes very important  in the
analysis of the final states to be given in the next section.

As far as the total lifetime is concerned, the most relevant effect
is produced by WA which to lowest order in the strong
interactions leads to a moderate reduction in lifetime:
$$\tau (B_c) \sim 4 \cdot 10^{-13}\; sec \; . \eqno(23)$$

\section{On the Pattern in the Final States}

As stated already in eq.(1) PI reduces
$\Gamma _{b\ra c \bar cs}(B_c)$ considerably.
The interference between WA
and $b$ decays sketched above also reduces the charm content in the
final state of $\Delta B=1$ $B_c$ decays in general. The argument
goes a follows: To describe the impact of WA on the decay rate
beyond the lowest order in the strong interactions one has to
include the emission of `off-shell' as well as `on-shell' gluons
with the former hadronizing into a $c \bar c$ pair:
$$b \bar c \ra d\bar u/s\bar c + g^*g^*\ra
d\bar u/s\bar c + c \bar c$$
For this reaction can interfere with the lowest order decay
process. As shown in ref.\cite{MIRAGE}, this interference
is destructive and it actually will reduce
the rate for $B_c \ra s \bar c c \bar c$ and quite possibly also
for $B_c \ra d \bar u s \bar c$ by roughly 5-10\%.
This decrease in the decay
rate is largely compensated for by
$B_c \ra d\bar u/s\bar c + gg$ where the gluon hadronizes mainly
into {\em light-flavour} hadrons. Details will be discussed in a
future publication.

\section{Conclusions}

$B_c$ decays represent a particularly intriguing lab to study the
interplay of strong and weak forces in a non-trivial environment.
The $1/m_Q$ expansion derived from QCD makes clear
predictions on the global pattern:

\noindent $\bullet$ a short $B_c$ lifetime well below 1 psec;

\noindent $\bullet$ a preponderance of charm over beauty
decays among the non-leptonic modes;  and

\noindent $\bullet$ a reduced semileptonic branching ratio
with roughly equal contributions from $b\ra c l \nu$ and
$c \ra s l \nu$.

\noindent Essential for the analysis is the observation that
in a treatment genuinely based on QCD there can be {\em no}
corrections of order $1/m_Q$ that have been introduced in
purely phenomenological models and play a central role there.

As far as the numerical predictions are concerned, one has to keep
an important caveat in mind: the weak link in the analysis is the
fact that the charm quark mass does not provide a parameter that
is very large compared to ordinary hadronic scales. Thus the
$1/m_c$ expansion cannot be expected to be quickly convergent.
In principle it is conceivable that it might actually fail in
charm transitions, say through quark-hadron duality
becoming inoperational there \cite{MISHA}.

{\bf Acknowledgements:}  This work was
supported by the National Science Foundation under
grant number PHY 92-13313.


\begin{thebibliography}{99}
\bibitem{QUIGG}
E. Eichten, C. Quigg, {\em Phys. Rev.} {\bf D49} (1994) 5845.

\bibitem{LUSIGNOLI}
M. Lusignoli, M. Masetti, {\em Z. Physik} {\bf C51} (1991) 549.

\bibitem{SV}
for the first suggestion, see: M. Shifman, M. Voloshin, 1982, in: V.
Khoze, M. Shifman, {\em Uspekhi Fiz. Nauk} {\bf 140} (1983) 3
[{\em Sov. Phys. Uspekhi} (1983) 387]; {\em Sov. Journ. Nucl. Phys.}
{\bf 41} (1985) 120.

\bibitem{BUV}
I.I. Bigi, N.G. Uraltsev, A. Vainshtein, {\em Phys. Lett.} {\bf B293}
(1992) 430; (E) {\bf B297} (1993) 477;
B. Blok, M. Shifman, {\em Nucl. Phys.} {\bf B399} (1993) 441; 459.

\bibitem{CGG}
J. Chay, H. Georgi, B. Grinstein,
{\em Phys. Lett.} {\bf B247} (1990) 399.

\bibitem{OPTICAL}
I.I. Bigi, M. Shifman, N.G. Uraltsev, A. Vainshtein,
{\em Phys. Rev.} {\bf D} (1995) .

\bibitem{PETRARCA}
G. Altarelli, S. Petrarca,
{\em Phys. Lett.} {\bf B261} (1991) 303.

\bibitem{MIRAGE}
I.I. Bigi, N.G. Uraltsev, {\em Phys. Lett.} {\bf B280} (1992) 120;
{\em Nucl. Phys.} {\bf B423} (1994) 33; the detailed discussion
given in these papers for $B_{u,d,s}$ decays is easily carried over
to $B_c$ decays.

\bibitem{MISHA}
For a discussion of these issues see: M. Shifman, talk given
at the V International Symposium on Particles, Strings
and Cosmology -- PASCOS --, John Hopkins University, Baltimore,
March 1995, to appear in the Proceedings, preprint
TPI-MINN-95/15-T [hep-ph/9505289].












\end{thebibliography}
\end{document}